\documentstyle[pre,aps,epsfig,multicol]{revtex}

\newcommand{\anfanglinie}{{\hfill\rule[-1em]{0.1pt}{1em}\rule{3.5in}{0.1pt}}}
\newcommand{\schlusslinie}{{\noindent\rule{3.5in}{0.1pt}\rule{0.1pt}{1em}}}
\newcommand{\mycaption}[1]{\caption{\narrowtext #1}}

\newcommand{\vek}[1]{{\mathbf #1}}
\newcommand{\interf}{{\cal I}}

\begin{document}
\draft

\title{Porous silicon formation and electropolishing} 

\author{Markus Rauscher\footnote{email: rauscher@ccmr.cornell.edu}}
\address{Laboratory of Atomic and Solid State Physics (LASSP), 
Clark Hall, Cornell University, Ithaca, NY 14853, USA}

\author{Herbert Spohn}
\address{Zentrum Mathematik und Physik Department, 
TU M\"{u}nchen, D-80290 M\"{u}nchen, Germany}

\date{\today}

\maketitle

\begin{abstract}
Electrochemical etching of silicon in hydrofluoride containing electrolytes 
leads to pore formation for low and to electropolishing for
high applied current. The transition between pore formation and polishing is
accompanied by a change of the valence of the electrochemical dissolution 
reaction.
The local etching rate at the interface between the semiconductor and the
electrolyte is determined by the local current density. We model the transport
of reactants and reaction products and thus the current density in both, the 
semiconductor and the electrolyte. Basic features of the chemical reaction at
the interface are summarized in law of mass action type boundary conditions for
the transport equations at the interface.
We investigate the linear stability of a planar and flat 
interface. Upon increasing 
the current density the stability flips either through a change of the valence of the
dissolution reaction or by a nonlinear boundary conditions at the interface.
\end{abstract}


\begin{multicols}{2}


\tableofcontents

\section{Introduction}

\label{introduction}
Porous silicon was discovered in the fifties trying to electropolish
silicon in hydrofluoric acid \cite{ulhir,turner}\/. For low current 
densities,
respectively high electrolyte concentrations, silicon is not electropolished 
but pores are
formed. Increasing the current density over a threshold value, which  decreases
with the electrolyte concentration, results in electropolishing.
In the beginning of the
nineties visible luminescence at room temperature was discovered
\cite{canham,lehmann}\/. The possibility to produce optoelectronic devices out
of porous silicon started enormous research activity. Meanwhile many
applications for porous silicon are in development. Most of these applications
are based on the morphology of porous silicon, for a review see \cite{parkhutik99}\/. 

Porous silicon is formed by anodic dissolution of silicon in hydrofluoric
acid. The silicon surface is in contact with the electrolyte, usually in a
Teflon cell. Through the external potential an electric current is 
maintained across the cell and flows  from the
semiconductor to the acid.  Defect electrons (i.e.\/ holes) from the
semiconductor and HF (or F$^-$ ions) from the electrolyte combine at the
fluid-semiconductor interface and dissolve silicon through an  electrochemical 
reaction. The morphology 
of the unsolved silicon depends on the current. In the 
electropolishing phase the silicon surface is etched 
layer by layer and remains essentially 
flat, whereas in 
the porous silicon phase many holes are formed of a size ranging 
from a few nanometers to microns. Porosities of over 95\% relative to crystalline
silicon can be reached.

Despite its importance, there is little theoretical understanding of how
porous silicon is formed, for a review see \cite{john}\/. 
Even the basic issue why there is a transition
from porous silicon formation to electropolishing is unresolved. The
reasons are rather obvious, when one compares with other
pattern formation processes like dendrites, viscous fingering, and colloidal
aggregation. As in these systems, we have to understand the dynamics of a
moving interface, here between a semiconductor and an acid. At the interface
silicon is dissolved through an electrochemical reaction. Thus in 
contrast to the better understood systems mentioned above, we
have to consider the transport of several species in the presence
of an electric field. Moreover, since the
species feed back into the electrical field and react with each other,
the transport equations become nonlinear. 

One approach to model the dissolution process is by a stochastic growth 
model, i.e.\/ growth of the fluid into the semiconductor, see for example 
\cite{erlebacher94,collins89}\/. These models are inspired by the diffusion
limited aggregation (DLA) model. While, in principle, such a model is on the
atomic scale, because of numerical limitations in practice larger spatial
units are used. By suitable adjustments of model parameters structures
qualitatively similar to porous silicon can be produced, but the quantitative
connection of model parameters to physical parameters is lost and in some
cases the coarsening introduces a new length scale into the system 
which obscures the
physics.

A second approach, starting at much larger length scales, is to use a
continuum description for both, the motion of the interface and the ionic
and electronic transport, see e.g., \cite{parkhutik93,kang93,valance97,foell00}\/.
These models include a depletion or passivation layer at the interface
in a phenomenological way and assume an ad hoc surface tension to stabilize
against small perturbations and to provide a length scale which can be
compared with the pore formation. However, surface tension can only affect
length scales of the order of the micropore diameter, i.e.\/ nanometers. 
The above mentioned models are not valid on these small scales since 
the mean free path, quantum effects, and the electrical double layer in the 
electrolyte would have to be taken into account.

We will use here also a description through continuum equations, but take
care to model the actual chemistry and kinetics at the interface and the 
physical transport mechanisms. This results in a somewhat complicated 
set of evolution equations and we have to be satisfied with the 
more modest goal
to understand whether continuum equations in general are able to predict
the transition from electropolishing to pore formation. As benchmark for the
transition we use the dispersion relation as obtained from a 
linear stability analysis of the flat
interface moving at a constant velocity. If the dispersion switches from
unstable to stable, we interpret this as the transition from pore
formation to polishing. Of course, if the continuum equations contain the
information of the pore structure at all, it will not be unraveled in 
such a stability analysis.

In Section~\ref{modeling} we discuss the full nonlinear transport equations 
and their boundary conditions at the interface. We derive a simplified 
transport model which covers essential features of electrochemical etching of
semiconductors and calculate the stability of the 
dissolution front in linear order in
Section~\ref{stabilityanal}\/. Our results are summarized in
Section~\ref{discussion}\/.

\section{Modeling electrochemical etching}
\label{modeling}

\subsection{Electrochemistry}
While anodizing silicon in hydrofluoric acid, silicon is
dissolved in an electrochemical reaction. The detailed reaction mechanism is
still topic of actual research. However, during pore formation hydrogen
evolution is observed whereas no hydrogen is formed during electropolishing. 
The valence of the chemical reaction differs in both cases. During pore formation the dissolution steps for a single
silicon atom add up to \cite{gerischer}
\begin{equation}
\text{Si} + 6\, \text{H\/F} +
\text{h}^+ \rightleftharpoons
\text{Si\/F}^{2-}_6 + \text{H}_2 + 4\, \text{H}^+ + 
\text{e}^-.
\label{porereaction}
\end{equation}
For each silicon atom two elementary charges have to reach the interface. 
For electropolishing the sum reaction is \cite{lehmanndiss}
\begin{equation}
\text{Si} + 6 \text{H\/F} + 4\, \text{h}^+
\rightleftharpoons
\text{Si\/F}^{2-}_6 + 6 \text{H}^+.
\label{polierreaction}
\end{equation}

In both cases, for pore formation and for electropolishing, the local
silicon dissolution rate is proportional to the local electrical 
current density component
normal to the interface, denoted by $\vek{j}_\perp$. Thus the local interface 
velocity $\vek{w}$ is
\begin{equation}
\vek{w}=-F \, \vek{j}_\perp,
\end{equation}
where $F$ is the volume of silicon per unit area dissolved by a unit charge.
This number is inversely proportional to the valence $\nu$ of the chemical
reaction, i.e.\  the number of unit charges necessary to dissolve one silicon
atom. The valence doubles as the current density is increased over the
threshold value for electropolishing \cite{lehmanndiss}\/.

\subsection{Transport equations}
The local current density is determined by the transport of reactants and
reaction products in the semiconductor and the electrolyte as well as by the
reaction kinetics. Modeling the transport and the interface reaction depends
strongly on the considered length scale. In n-doped silicon, the pore spacing
is typically some microns whereas in p-silicon nanometer sized pores are
formed. 

The shorter the considered length scale the more detailed the model has to be.
The mean free path of charge carriers in the semiconductor is of the
order of some ten nanometers. Transport on this length scale can be described 
by Boltzmann equations, but their nonlocality makes them difficult to analyze 
\cite{cercignani88}\/. 

On the nanometer scale, quantum effects start to
play a role \cite{lehmann} and the electrical double layer at the 
interface in the electrolyte, i.e.\/ the Helmholtz layer, 
has to be taken into account \cite{bockris}\/. At this level, details 
about the electrochemical reaction path way have to be fed into electronic 
structure calculations to determine the boundary conditions. Such detailed knowledge 
is not available and the presence of an electrolyte makes the calculations even more 
complicated.

To avoid these difficulties we restrict our model to length scales
large as compared to the mean free path in the semiconductor,
i.e.\ larger than 100~nm\/. Then, the transport in the semiconductor and in the
electrolyte can be described by drift and diffusion, i.e.\ by 
Nernst-Planck
equations. The current density of electrons, $\vek{j}_n$, and holes, 
$\vek{j}_p$,
is given by 
\begin{eqnarray}
\label{seminernstplanck}
\vek{j}_n &=&\phantom{-} e\,D_n\,\vek{\nabla}n+ 
e^2\,\mu_n\,n\,\vek{E},\\
\vek{j}_p&=&-e\,D_p\,\vek{\nabla}p + e^2\,\mu_n\,p\,\vek{E}\nonumber,
\end{eqnarray}
where $D_{n/p}$ is the diffusion constant, $\mu_{n/p}$ the mobility, $e$ 
the elementary charge, and $\vek{E}$  the electric field. The electron and
hole concentrations $n$ and $p$ determine the local charge density and thus 
the electric field via the Poisson equation
\begin{equation}
\vek{\nabla}\cdot \vek{E} = \frac{e}{\epsilon_{\text{Si}}}\,(p-n+N),
\end{equation}
with $N$ the density of ionized dopant atoms and $\epsilon_{\text{Si}}$ 
the dielectric constant of silicon. For p-doped silicon $N$ is
negative and for n-doped silicon positive. Since the electric field is
determined by the charge carrier concentration, the products $n\,\vek{E}$ and
$p\,\vek{E}$ in Eq.~(\ref{seminernstplanck}) represent nonlinearities.
Another nonlinearity appears due to production and recombination of electron
hole pairs in the continuity equation 
\begin{equation}
\label{eleholepair}
\vek{\nabla}\cdot\vek{j}_n = \frac{e}{\tau}\, \frac{p\, n -
p_{\text{eq}}
n_{\text{eq}}}{p_{\text{eq}}+n_{\text{eq}}}=-\vek{\nabla}\cdot\vek{j}_p,
\end{equation}
where $\tau$ the life time of the charge carriers and  $n_{\text{eq}}$ 
and $p_{\text{eq}}$ is the equilibrium electron and
hole concentration, respectively
\cite{henisch}\/. We make the quasistatic approximation neglecting the time
derivative of the concentration fields. This is justified if the relaxation
time for any field involved in the dissolution process is much faster than
the interface movement.
The electrical charge is conserved due to $\vek{\nabla}\cdot
\vek{j} = \vek{\nabla}\cdot(\vek{j}_n + \vek{j}_p) =0$\/.
The source term in Eq.~(\ref{eleholepair}) is derived from a law of mass
action for the recombination reaction 
$\text{e}^-  + \text{h}^+ \rightleftharpoons 0$\/.

If convection is negligible,
the transport of molecules and ions in the electrolyte can be described 
analogously. $\text{HF}$, $\text{H}_2$, $\text{H}^+$ and $\text{SiF}_6^{2-}$ 
have to be considered as well as fluoride $\text{F}^-$ and $\text{OH}^-$\/.
For each component $\text{X}$ a Nernst-Planck equation gives the particle
current density, denoted by $\vek{i}_{\text{X}}$ (in contrast to the electric
current density $\vek{j}$),
\begin{equation}
\label{nernstplanck}
\vek{i}_{\text{X}} = - \, D_{\text{X}}\,\vek{\nabla} C_{\text{X}}
+q_{\text{X}}\,\mu_{\text{X}}\,C_{\text{X}}\,\vek{E},
\end{equation}
with the particle's charge $q_{\text{x}}$\/. The reactions $\text{HF}
\rightleftharpoons \text{H}^+ + \text{F}^-$ and $\text{H}_2\text{0}
\rightleftharpoons \text{H}^+ + \text{OH}^-$ have to be taken into account in
source terms for the continuity equations of the corresponding current
densities 
\begin{eqnarray}\label{current}
\vek{\nabla}\cdot\vek{i}_{\text{OH}^-}&=&\frac{1}{\tau_{\text{H}_2\text{0}}}\, 
\frac{C_{\text{OH}^-}\,C_{\text{H}^+}- K_w}{C_{\text{H}_2\text{O}}^2}\,, 
\nonumber\\
\vek{\nabla}\cdot\vek{i}_{\text{H}^+}&=&
-\frac{1}{\tau_{\text{H}_2\text{0}}}\,
\frac{C_{\text{OH}^-}\,C_{\text{H}^+}- K_w}{C_{\text{H}_2\text{O}}^2} \nonumber\\
&& -\frac{1}{\tau_{\text{HF}}}\,
\frac{\frac{C_{\text{F}^-}\,C_{\text{H}^+}}{C_{\text{HF}}}-
K_{\text{HF}}}{C_{\text{H}_2\text{O}}}\,, \\
\vek{\nabla}\cdot\vek{i}_{\text{F}^-}&=&
\frac{1}{\tau_{\text{HF}}}\,
\frac{\frac{C_{\text{F}^-}\,C_{\text{H}^+}}{C_{\text{HF}}}-
K_{\text{HF}}}{C_{\text{H}_2\text{O}}} \nonumber.
\end{eqnarray}
The equilibrium constants for the water dissociation and the HF hydration
are $K_w=10^{-14}\,\frac{\text{mol}^2}{\ell^2}$ and
$K_{\text{HF}}=3.5\,10^{-4}\,\frac{\text{mol}}{\ell}$ \cite{atkins86}\/.
In (\ref{current}) the water concentration, considered as a natural 
constant of unit $\frac{\text{mol}}{\ell}$, 
has been added to correct the units \cite{remarkonunits}\/. 
In the electrolyte the Poisson equation is
\begin{equation}
\label{poissoneq}
\vek{\nabla}\cdot\vek{E} = \frac{e}{\epsilon_{\text{Ele}}}\,
\big(
C_{\text{H}^+}-C_{\text{F}^-}-C_{\text{OH}^-}-2\,C_{\text{SiF}_6^{2-}}
\big),
\end{equation}
with the electrolyte's dielectric constant $\epsilon_{\text{Ele}}$\/.
As in the semiconductor, the law of mass action type source terms and the
coupling of the ion concentrations to the electric field represent
nonlinearities which make the transport equations considerably more
complicated as compared to equations used to describe directed
solidification or viscous fingering for example \cite{pelce}\/. Moreover,
especially in the description of the transport in the electrolyte, there are
poorly understood features. First of all, there are many other ions in the
electrolyte that do not take part in the dissolution reaction
directly but affect the electric field and the transport properties. 
Secondly,
during pore formation, silicon enters the solution as $\text{HSiF}_{3}$
and reacts to $\text{SiF}_6^{2-}$ in the solution \cite{gerischer}\/. The
reaction rate for this process is not known and it is possible that 
a considerable amount of $\text{HSiF}_{3}$ is present near the
interface in 
solution. Thirdly, hydrogen bubble evolution is not modeled as well as
convection. Nevertheless, we propose to proceed with the nonlinear 
transport equations (\ref{seminernstplanck}) to (\ref{poissoneq})\/.

\subsection{Boundary conditions}
At the interface between the semiconductor and the electrolyte, the
dissolution reactions Eqs.~(\ref{porereaction}) and (\ref{polierreaction})
have to be taken into account. In both equations, the electric current density
component normal to the interface (denoted by the subscript $\perp$)
is given by the difference of the forward
and backward reaction rate. In a law of mass action type approximation,
these rates are given by the concentration product of the reactants and the
reaction products, respectively. Thus for the pore formation reaction we
obtain 
\begin{equation}
\label{bcpore}
j_\perp = -\Gamma_{\text{pore}}\, \big(
C_{\text{HF}}^6\,p-\eta_{\text{pore}}\,C_{\text{SiF}_6^{2-}}\, C_{\text{H}_2}\,
C_{\text{H}^+}^4\, n
\big),
\end{equation}
with the reaction rate $\Gamma_{\text{pore}}$\/. The parameter 
$\eta_{\text{pore}}$ is a measure for the
equilibrium concentration product for this reaction. 
The current density is taken positive
for currents flowing from the electrolyte into the semiconductor. For reaction
Eq.~(\ref{polierreaction}) we obtain correspondingly
\begin{equation}
\label{bcpolish}
j_\perp = -\Gamma_{\text{polish}} \, \big(
C_{\text{HF}}^6\,p^4 - \eta_{\text{polish}}\, C_{\text{SiF}_6^{2-}}\,
C_{\text{H}^+}^6 
\big).
\end{equation}
In both cases, the concentration of crystalline silicon is a constant 
summarized in the parameters $\Gamma$ and $\eta$\/. 
Surface tension could be accounted for by a curvature dependence of $\eta$
but cannot play a role at the length scales discussed here.
Obviously, these boundary conditions are
highly nonlinear. The particle current density components normal to the
interface 
represent the stoichiometry of the corresponding chemical
reaction. In case of Eq.~(\ref{porereaction}) this is
\begin{eqnarray}
4\,{i_{\text{HF}}}_\perp &=& 6\,{i_{\text{H}^+}}_\perp\,,\nonumber\\
{j_p}_\perp &=& {j_n}_\perp\,,\nonumber\\
\frac{1}{e}\,{j_p}_\perp &=& {i_{\text{H}_2}}_\perp\,,\label{porecurrent}\\
\frac{4}{e}\,{j_p}_\perp &=& {i_{\text{H}^+}}_\perp\,,\nonumber\\
\frac{1}{e}\,{j_p}_\perp &=& {i_{\text{SiF}_6^{2-}}}_\perp\,,\nonumber
\end{eqnarray}
and for Eq.~(\ref{polierreaction})
\begin{eqnarray}
{i_{\text{HF}}}_\perp &=& {i_{\text{H}^+}}_\perp\,,\nonumber\\
\frac{1}{e}\,{j_p}_\perp &=& 4\,{i_{\text{SiF}_6^{2-}}}_\perp\,,\label{policurrent}\\
\frac{6}{e}\,{j_p}_\perp &=& 4\,{i_{\text{H}^+}}_\perp.\nonumber
\end{eqnarray}
Thus the current densities of all species taking part in the reaction are
determined by fixing the current density of one of them. The normal current
density component of species which do not take part in the dissolution reaction
is zero.

The inner Helmholtz layer in the electrolyte is not described by the ion
transport equations. This layer has a very high capacity as compared to the 
diffuse part of the electrical double layer in the electrolyte and the depletion
layer in the semiconductor \cite{vanmaekelbergh}\/. For this reason, the main
potential drop across the interface occurs in the depletion layer in the
semiconductor. Thus the electrical potential $V$
($\vek{E}=-\vek{\nabla}V$) is continuous at the interface
in a first approximation.
For the electric field boundary conditions as known from electrostatics are used,
i.e.\ the tangential component of $\vek{E}$ and the normal component of
$\epsilon\,\vek{E}$ are continuous across the interface.

We do not specify the boundary conditions at the cathode or at the backside of
the wafer, since no experimental evidence is reported for an influence of the
cathode or the wafer backside on the dissolution process. 
Outside the depletion layer the semiconductor
is electrically neutral and can be treated as an
Ohmic conductor. The electron and hole concentrations have equilibrium values.
In the electrolyte, there is more variety. Due to the
consumption of reactants and the accumulation of reaction products, the
composition of the electrolyte changes in time. However, usually the 
electrolyte is stirred. In a large container, this means, that in a certain
distance from the anode, the electrolyte is homogeneous, approximately in
equilibrium, and has a composition which hardly changes in time. Thus, a
reasonable boundary condition for the model equations is to fix the
concentration of the electrolyte components at a certain distance from the
interface to the equilibrium values. This distance depends strongly on stirring
and the current density and must remain as a free parameter. Such kind of
boundary condition is a very crude approximation, since convection certainly
plays a role even in the diffusion layer, i.e.\ the region near the interface
where the electrolyte is not homogeneous due to the applied current. To include
convection in the model is in principle possible.
On the scale of the pores convection should
not play a  role because of the high viscosity of hydrofluoric acid.

The model described in this section should be 
reasonably close to the physics and chemistry
of the etching process on length scales large compared to the mean free path
in the semiconductor. We have to deal with nonlinear transport equations and boundary
conditions. However, 
the two cases, i.e.\ pore formation and electropolishing, are actually treated 
as two separate models. One would like to have a model that decides itself
which reaction pathway (i.e.\/ which law of mass action boundary condition) to take, 
depending on concentrations or current densities at the interface.
Combining the two boundary conditions
Eq.~(\ref{bcpore}) and Eq.~(\ref{bcpolish}) to 
\begin{eqnarray}
\label{combinedmwg}
j_\perp& =& -\Gamma_{\text{pore}}\, \big(
C_{\text{HF}}^6\,p-\eta_{\text{pore}}\,C_{\text{SiF}_6^{2-}}\, C_{\text{H}_2}\,
C_{\text{H}^+}^4\, n\big)\nonumber\\
&&-\Gamma_{\text{polish}} \, \big( 
C_{\text{HF}}^6\,p^4 - \eta_{\text{polish}}\, C_{\text{SiF}_6^{2-}}\,
C_{\text{H}^+}^6 
\big)
\end{eqnarray}
gives the correct normal current density. But the stoichiometry equations 
(\ref{porecurrent}) and (\ref{policurrent}) have to be included too. For that one
needs to know the local fraction of silicon atoms that are dissolved by the 
pore reaction (\ref{porereaction}) and the polishing reaction 
(\ref{polierreaction}), respectively. 

Beside this constraint, analyzing the above developed model (analytically 
or numerically) would be a really demanding venture. The parameter space 
is large and even though many parameters have a direct physical interpretation,
experimental values are not available to fix them. 

Our goal is to study the transition from pore formation to electropolishing,
in particular to investigate the mechanism that can lead to such a transition. 
Therefore we study a simplified model that captures key features of the above
description of electrochemical dissolution of silicon, namely the interplay of more 
than one field determining the interface motion, a change of valence of the
dissolution reaction, and law of mass action type boundary condition.

\section{Linear stability analysis}

\label{stabilityanal}
\subsection{Simplified model}
\label{simplmodel}
\subsubsection{Model equations}
To have a tractable model and to gain some experience we have to simplify 
and consider only one field, $\Psi_E$, in the
electrolyte and one field, $\Psi_S$, in the semiconductor. These fields
could  be either the concentrations of one of the species in the
electrochemical reaction, or the electrical potential, or a linear combination
of fields as, e.g., the total amount of fluor per unit volume which is the sum
$C_{\text{F}^-} + C_{\text{HF}}$\/. For simplicity, we will work
with concentration fields in the following.
 To account 
for the interaction between the various fields we
include source terms in the continuity
equations which drive the fields to their equilibrium values
$\Psi_E^{\text{eq}}$ and $\Psi_S^{\text{eq}}$ in the electrolyte and
the semiconductor, respectively. We assume only diffusive transport
to keep the transport equations linear, which allows the stability
analysis of a planar interface to be performed analytically. 
Nonlinear transport equations would lead to linearized equations 
for a flat and planar interface with nonconstant coefficients. The
particle current densities in the electrolyte and the semiconductor,
$\vek{i}_E$ and $\vek{i}_S$, respectively, are assumed to be  given by
\begin{equation}
\vek{i}_E = -D_E\, \vek{\nabla}\Psi_E 
\quad\text{and}\quad
\vek{i}_S = -D_S\, \vek{\nabla}\Psi_S,
\end{equation}
with the diffusion constants $D_E$ and $D_S$\/. The continuity
equations are then 
\begin{equation}
\label{continuity}
\vek{\nabla}\cdot \vek{i}_{E/S} = - D_{E/S} \,
\vek{\nabla}^2\,\Psi_{E/S} = 
-\frac{1}{\tau_{E/S}}\, (\Psi_{E/S}-\Psi_{E/S}^{\text{eq}}),
\end{equation}
with the time constants $\tau_E$ and $\tau_S$ for the electrolyte
and the semiconductor, respectively \cite{voltage}\/.  

At the moving interface $\interf= \left\{(x,y,z)\,|\, z=h(x,y;t)
\right\}$ we assume a law of mass action type boundary condition
for the reaction 
\begin{equation}
\label{interfreact}
\Psi_E + \Psi_S \rightleftharpoons 0,
\end{equation}
similar to 
Eq.~(\ref{bcpore}) or (\ref{bcpolish})\/. The setup is illustrated in
Fig.~\ref{setupfig}\/. Since we want to perform a
linear stability analysis of a flat and planar interface the
restriction to single valued interfaces is no real limitation for
our analysis. 
Another choice of the interface reaction would be $\Psi_E
\rightleftharpoons \Psi_S$, leading to linear law of mass type
boundary conditions, see \cite{rauscher00}\/. This type of boundary
conditions does not lead to the desired properties.
The law of mass action for Eq.~(\ref{interfreact}) is 
\begin{equation}
\label{massaction}
{i_S}_\perp \big|_\interf = - \Gamma\,\big(\Psi_S\,\Psi_E
\big|_\interf -\eta\big)
\end{equation}
and the stoichiometry is represented by
\begin{equation}
\label{soichiometry}
{i_S}_\perp \big|_\interf = -{i_E}_\perp \big|_\interf.
\end{equation}

At some distance $d_E$ and $d_S$ from the interface $\interf$,
either the current density $\vek{i}_{E/S}$ or the field $\Psi_{E/S}$
can be fixed. 
However, at least at one side the field has to be fixed to eliminate all
gauge freedom. We will fix the current density in the semiconductor and
the field in the electrolyte, i.e.\/ ${i_S}|_{z=d_s} = I$ and
$\Psi_E|_{z=-d_E} = \Psi_{d_E}$\/.
The normal velocity $\vek{w}$ of the interface is given by the
normal current density 
\begin{equation}
\vek{w} = -F\,{i_S}_\perp\big|_\interf.
\end{equation}
With the choice of sign in the above equation and interface reaction
(\ref{interfreact}) the particle current in the semiconductor has to flow
to the interface to dissolve the silicon. This means that $\Psi_S$ 
has to be the hole concentration.

A change in the valence of the dissolution reaction can then be
modeled by a current dependent $F({i_S}_\perp)$, which is inverse
proportional to the valence $\nu\propto \frac{1}{F}$, i.e.\ the
number of particles $\Psi_{E/S}$ needed to dissolve a certain amount
of semiconductor material. Geometric considerations lead to the
growth rate of the height function $h(x,y;t)$ 
\begin{equation}
\label{interfvelo}
\dot{h}(x,y;t)= -F({i_S}_\perp)\,\left(
\begin{array}{c} -\vek{\nabla} h \\ 1 \end{array}\right)\cdot
{\vek{i}_S} \bigg|_\interf ,
\end{equation}
where $\dot{h}$ denotes the time derivative of $h$\/.

\subsubsection{Linearized theory}
For a flat and planar interface $h_0(t)$, the transport equations
(\ref{continuity}) become ordinary linear
differential equations with constant coefficients in the independent
variable $z$\/. The solutions, which can be obtained analytically,
depend on the boundary conditions at $z=d_E$ and $z=d_S$ and will be
discussed later. They are linear combinations of exponentials or, in the
limiting case $\tau_{E/S}\to\infty$, affine functions of $z$\/.

Now we assume a small perturbation $\delta h(x,y;t)$ of a certain
wave length $\lambda=\frac{2\,\pi}{k}$ of the interface and expand
the fields (and correspondingly the current densities) up to 
first order in $\delta h$,
\begin{eqnarray}
\Psi_{E/S} & = & \Psi_{E/S}^0 + \delta \Psi_{E/S} + {\cal O}(\delta
h^2),\\
\vek{i}_{E/S} & = & \vek{i}_{E/S}^0 + \delta \vek{i}_{E/S} + {\cal
O}(\delta h^2).
\end{eqnarray}
From Eq.~(\ref{interfvelo}) the time evolution of the perturbation
can be derived
\begin{equation}
\label{perturbevol}
\delta\dot{h} = -F({i_S^0}_\perp)\,\left(
\delta h \, {i_S^0}'_\perp + \delta{i_S}_\perp 
\right)\,
\underbrace{\left(1+{i_S^0}_\perp\,\frac{{\rm d} \ln F}{{\rm d}
{i_S}_\perp}\Big|_{{i_S^0}_\perp}\right)}_{(*)}
\Bigg|_\interf,
\end{equation}
with the prime abbreviating the derivative with respect to
$z$\/.
The last term $(*)$ is independent of the shape of the perturbation
$\delta h$\/. In terms of the valence $\nu$ it can be written as 
\begin{equation}
(*)=\left(1-{i^0_S}_\perp\,\frac{{\rm d} \ln \nu}{{\rm d}
{i_S}_\perp}\Big|_{{i^0_S}_\perp}\right). 
\end{equation}

The growth speed of the perturbation $\delta h$ is thus proportional to
$\delta h$ and Eq.~(\ref{perturbevol}) can be written as $\delta\dot{h} =
\omega(k)\, \delta h$ with the dispersion relation $\omega(k)$\/.
Perturbations with $\omega(k)>0$ will grow exponentially and are called
unstable, whereas modes with $\omega(k)<0$ are damped and stable.

From the continuity equations (\ref{continuity}) it
follows 
\begin{equation}
\label{firstorder}
-D_{E/S}\,\left(\delta\Psi_{E/S}'' - k^2 \, \delta
\Psi_{E/S}\right) = -\frac{1}{\tau_{E/S}}\,\delta \Psi_E
\end{equation}
 by comparing powers of $\delta h$. The first order terms in the boundary conditions
Eq.~(\ref{massaction}) and (\ref{soichiometry}) for the perturbed
fields $\delta\Psi_{E/S}$ are 
\end{multicols}\schlusslinie
\begin{equation}
\left(\delta {i_S}_\perp +\delta h\,
{i_S^0}'_\perp\right)\Big|_\interf =
-\Gamma\,\bigg(\big(\Psi_S^0 +
\Psi_S^{\text{eq}}\big)\,\left(\delta\Psi_E + \delta h\,
{\Psi_E^0}'\right)+
\big(\Psi_E^0 + \Psi_E^{\text{eq}}\big)\,\left(\delta \Psi_S + 
\delta h \, {\Psi_S^0}'\right)\bigg)\bigg|_\interf, 
\end{equation}
\anfanglinie\begin{multicols}{2}

\begin{equation}
\left(\delta {i_S}_\perp +\delta h\,
{i_S^0}'_\perp\right)\Big|_\interf =
-\left(\delta {i_E}_\perp +\delta h\,
{i_E^0}'_\perp\right)\Big|_\interf.
\end{equation}
At the lines $z=d_E$ and $z=d_S$ the fields satisfy 
Dirichlet boundary conditions.
 
\subsection{Change of valence}

Independent of the solution of the first order equation (\ref{firstorder}),
the third term $(*)$ in the right hand side of the time evolution equation
(\ref{perturbevol}) changes sign with the current density $i^0_{S_\perp}
 |_{\cal I}$ passing through the interface if $F$ (or the valence $\nu$)
varies strongly enough. A change of sign of this term changes the sign of
$\omega(k)$ for all $k$, making stable modes unstable and vice versa. A
similar mechanism has been proposed to explain the stability of the
macropore front \cite{lehmann93}\/.

A sharp change of the valence of the electrochemical dissolution reaction
from 2 to 4 electrons per Si atom at the critical current density for
electropolishing has been found experimentally. Estimating the valence
change from \cite[p.~39, Fig.~3.8]{lehmanndiss} leads to $i^0_{S_\perp}\,
\frac{d\ln \nu}{d i_{S_\perp}}\approx 1.18$, i.e.\/ enough to change the
sign of $\omega(k)$\/.

This change of sign leads only to a stabilization of the interface if there
have been no stable modes for lower current densities, since these would
become unstable. Moreover, our analysis cannot explain, why the interface
remains stable for high current densities, i.e.\/ for electropolishing,
where the valence does not change any more.

\subsection{Transport and boundary conditions}

The linearized interface growth model as described in
Sec.~\ref{simplmodel} can  in principle be solved analytically for all types
of boundary conditions far away from the interface. Among the many 
possibilities we
discuss some instructive limiting cases and infer the general behavior
from them. In this section  we assume for simplicity that $F$, and thus the 
valence $\nu$, is independent of the current density.

\subsubsection{Infinite life time---double Laplacian growth}

The simplest case is the limit of infinite life time $\tau_{E/S}\to
\infty$\/. Both fields, $\Psi_E$ and $\Psi_S$, solve the Laplace equation
and this simplified model is a straight forward extension of the well
studied Laplacian growth model \cite{pelce}\/.
The solutions for the planar interface are then 
\begin{eqnarray}
\Psi^0_S & = & -\frac{I}{D_S}\, z + \frac{D_E\, (-I + \Gamma\,
\eta)}{\Gamma\, (I\, d_E+D_E\,\Psi_{d_E}}, \nonumber\\
\Psi^0_E & = & \frac{I}{D_E}\, (z + d_E) + \Psi_{d_E}.
\end{eqnarray}

After solving the first order equations
(\ref{firstorder}) one obtains for the dispersion relation
\begin{equation}
\omega(k) = \frac{F\,I\,k\,\left(
\Psi^0_E(0)-
\frac{D_S}{D_E}\,\Psi^0_S(0)\right)}
{\Psi^0_E(0)\,\coth
k\,d_S +\frac{D_S}{D_E}\,\Psi^0_S(0)\,\tanh k\,d_E
+ \frac{D_S}{\Gamma}\,k}.
\end{equation}
For small $k$, the sign of $\omega(k)$ is basically determined by 
the ratio of the diffusivities and the current direction and goes quadratically
to zero
\begin{equation}
\omega(k) \sim F\,I\,\left( \frac{D_S\,\Psi^0_S(0)}{D_E\,\Psi^0_E(0)} -1
\right)\,d_E\,k^2 + {\cal O}(k^4).
\end{equation}
In the limit of $d_E\to \infty$ the dispersion relation goes linearly to zero
for $k\to 0$ with the same prefactor as above (i.e.\/ substitute $d_E\,k^2$ 
by $k$)\/. 

For large $k$, the dispersion relation saturates at 
\begin{equation}
\omega(k) \sim F\,I\,\Gamma\,\left(
\frac{D_S\,\Psi^0_S(0)}{D_E\,\Psi^0_E(0)} -1\right)\,
\frac{\Psi^0_E(0)}{D_S} + {\cal O}(\frac{1}{k}).
\end{equation}
Again the term $\left(\frac{D_S\,\Psi^0_S(0)}{D_E\,\Psi^0_E(0)} -1\right)$
determines the sign. For $D_E\gg D_S$ the dispersion relation is positive for
all $k$ but changes sign as $D_E\ll D_S$\/. In other words, the interface
is unstable if the front propagates into the medium with the much lower diffusion
constant. The ratio of diffusion constants at which the sign changes is 
determined by the other parameters, e.g., by $D_E/D_S=1/50$ in the example in
Fig.~\ref{doppelDLADfig}\/. The reason for the dependence of the stability on the
diffusion constant is the following. Reactants from the semiconductor (e.g., holes)
reach the pore tips first. The diffusive transport in the semiconductor thus
destabilizes the interface. On the other hand, the reactants in the electrolyte
reach the pore walls easier than the tips and thus stabilize the interface. 
If the
diffusion in the semiconductor is much slower than in the electrolyte, 
then the growth
speed is solely determined by the transport in the semiconductor and the interface
is unstable. This is the standard DLA scenario \cite{pelce}\/. 
In the opposite case, the transport
in the electrolyte determines the local dissolution rate and the interface is
linearly stable. This anti-DLA limit has been studied in \cite{DLEXX}\/.

The term $\left(\frac{D_S\,\Psi^0_S(0)}{D_E\,\Psi^0_E(0)} -1\right)$ also 
has two zeros as a function
of $I$ since the numerator is a  quadratic polynomial in $I$\/. 
For $D_E\gg D_S$ and a small negative $I$ the dispersion relation 
is positive for all $k$ and changes sign at the first zero $I_1$\/.
Fig.~\ref{doppelDLAIfig} illustrates this property of $\omega(k)$\/. This means
that the law of mass action boundary conditions provide a mechanism to
stabilize an interface simply by changing the current density. The second zero 
$I_2$ is smaller than 
$I_0=-\frac{D_E\,\Psi_{d_E}}{d_E}$ where $\Psi^0_E(0)$ changes sign. This
leads to an unphysical pole in the dispersion relation. Since we interpret 
the fields $\Psi_{E/S}$ as concentrations, $\Psi^0_E(0)$ has to be positive.

The finite distances $d_{E/S}$ have basically the effect that they provide
an infrared cutoff changing the dispersion relation from linearly to
quadratically for $k\to 0$\/ (besides changing the numerical values of
$\Psi^0_{E/S}$)\/. Therefore we set $d_E\to \infty$ for the discussion of the 
model with finite life times in the electrolyte.

\subsubsection{Helmholtz equation in the electrolyte}

To study the effect of a finite life time for the diffusing species, we take
$\tau_E$ finite while keeping $\tau_S$ infinite. The choice is motivated by
the high background ion concentrations in the electrolyte which provide a
buffer/reservoir for particles. Since we set $d_E\to \infty$, equilibrium
is the only choice for
the boundary condition in the electrolyte far away from the interface,
i.e.\/ $\Psi_E\to\Psi_E^{\text{eq}}$ for $z\to -\infty$\/.
 The solution for the flat interface is then
\begin{eqnarray}
\Psi^0_S(z) &=& -\frac{I}{D_S}\, z+
\frac{\eta-\frac{I}{\Gamma}}
{\frac{I}{D_E\,\kappa_E}+D_E\,\kappa_E\,\Psi^{\text{eq}}_E},
\nonumber\\
\Psi^0_E(z) &=& \frac{I}{D_E\,\kappa_E}\,{\rm
e}^{\kappa_E\,z}+\Psi^{\text{eq}}_E,
\end{eqnarray}
where $\kappa_E=1/\sqrt{D_E\,\tau_E}$ is the reciprocal diffusion length. 
The dispersion relation for this case is 
\end{multicols}\schlusslinie
\begin{equation}
\omega(k)= -F\,I\,k\,
\frac{\Gamma\,\left(\frac{I}{\kappa_E}+\Psi_E^{\text{eq}}\right)\,\sqrt{\kappa_E^2+k^2}-
D_S\,D_E\,\frac{\Gamma\,\eta-I}{\frac{I}{\kappa_E}+\Psi_E^{\text{eq}}}
\left(\sqrt{\kappa_E^2+k^2}-\kappa_E\right)}
{\left(D_E\,D_S\,k+\Gamma\,\left(\frac{I}{\kappa_E}+\Psi_E^{\text{eq}}\right)\,\coth
k\,d
\right)\, \sqrt{\kappa_E^2 +
k^2} + D_S\,\frac{\Gamma\,\eta-I}{\frac{I}{\kappa_E}+\Psi_E^{\text{eq}}}\, k}.
\end{equation}
\anfanglinie\begin{multicols}{2}

For small $k$, the sign of the dispersion relation is determined by the sign of
the current density $I$,
\begin{equation}
\omega(k)\sim -F\,I\,d_S\,k^2.
\end{equation}
For $d_S\to\infty$, $\omega(k)$  goes linearly to zero with the same
prefactor (i.e.\/ substitute $d_S\,k^2$ by $k$)\/. 

The dispersion relation has a limit for $k\to\infty$, which is a nonlinear
function of the current density $I$
\end{multicols}\schlusslinie
\begin{equation}
\label{dopdlafdlaoninf}
\omega(k)\sim-F\,I\,\frac{
\Gamma\,I^2+D_S\,\kappa_E\,(D_S\,\kappa_E+2\,\Gamma\,\Psi_E^{\text{eq}})\,I
+D_S\,\Gamma\,\kappa_E^2\,(D_S\,{\Psi_E^{\text{eq}}}^2 - D_S\,\eta)}
{D_S\,D_S\,\kappa_E\,(I+D_E\,\kappa_E\,\Psi_E^{\text{eq}})}
+{\cal O}(\frac{1}{k}).
\end{equation}
\anfanglinie\begin{multicols}{2}
The sign of this limiting value changes with the current density in the same way
as the sign of the dispersion relation in the previous section. The difference
here is, that the sign for small values of $k$ is independently fixed by the sign
of $I$\/. 
For $I<0$, i.e.\/ in the case of dissolution of the semiconductor, the
dispersion relation is positive for small $k$ and positive or negative for large
$k$\/. The change in the stability of small wave length modes is illustrated in
Fig.~\ref{dopDLAFDLAIfig}\/.

The reason why the sign of the dispersion relation for small $k$ is given by the
current density $I$ only is, that in this limit the electrolyte does not
influence the stability of the interface. The diffusion length $1/\kappa_E$
is a cutoff for the wave length up to which the electrolyte can stabilize the
interface. Longer wave length perturbations are controlled only by the
semiconductor. 

\subsubsection{Finite life time in semiconductor and electrolyte}

The findings in the above section lead to 
the conjecture that the stability of long
wave length perturbations is determined by the medium with the longer
diffusion
length. To verify this we assume a semi-infinite 
electrolyte and semiconductor to
keep formulae simple. With this type of equation and boundary condition it is no
longer possible to fix the current density $I$ but by changing the ration of
$\Psi_E^{\text{eq}}/\Psi_S^{\text{eq}}$ the current through the interface can be
controlled indirectly. The solutions for the flat interface are then
\begin{eqnarray}
\Psi_S^0(z)&=& B\, {\rm e}^{-\kappa_S\,z}+\Psi_S^{\text{eq}} \nonumber\\
\Psi_E^0(z)&=& \frac{D_S}{D_E\,\kappa_E}\, B\, {\rm e}^{\kappa_E\,z}+
\Psi_E^{\text{eq}},
\end{eqnarray}
where $B$ is the positive root of
\begin{eqnarray}
0&=&\Gamma\,D_S\,\kappa_S\,B^2 \nonumber\\
&+&\big(\Gamma\,(D_S\,\kappa_S\,\Psi_S^{\text{eq}} +
D_E\,\kappa_E\,\Psi_E^{\text{eq}}) + D_S\,D_E\,\kappa_S\,\kappa_E\big)\,
B\nonumber\\
&+& \Gamma\,D_E\,\kappa_E\, (\Psi_S^{\text{eq}}\,\Psi_E^{\text{eq}}-\eta).
\end{eqnarray}

We write the dispersion relation in terms of the zeroth order fields and current
density at the interface and we use the abbreviation ${\cal K}_{E/S} = 
\sqrt{\kappa_{E/S}^2+k^2}$
\end{multicols}\schlusslinie
\begin{equation}
\omega(k)=-F\,\Gamma\,{i_S^0}_\perp(0)\,\frac{
D_E\,{\cal K}_E\,({\cal K}_S-\kappa_S)\,\Psi_E^0(0) - 
D_S\,{\cal K}_S\,({\cal K}_E-\kappa_E)\,\Psi_S^0(0) }{
\Gamma\,\big(D_E\,{\cal K}_E\,\Psi_E^0(0) +D_S\,{\cal
K}_S\,\Psi_S^0(0)\big) + D_E\,D_S\,{\cal K}_E\,{\cal K}_S }.
\end{equation}

For long wave length, the dispersion relation is
\begin{equation}
\omega(k)\sim -F\,\Gamma\,{i_S^0}_\perp(0)\,\frac{\frac{1}{2\,\kappa_E\,\kappa_S} \,
(\kappa_E^2\,\Psi_E^0(0)-\kappa_S^2\,\Psi_S^0(0))}
{\Gamma\,\big(D_E\,\kappa_E\,\Psi_E^0(0)
+D_S\,\kappa_S\,\Psi_S^0(0)\big) + D_E\,D_S\,\kappa_E\,\kappa_S }\, k^2
+{\cal O}(k^4).
\end{equation}
\anfanglinie\begin{multicols}{2}
If the semiconductor has a much longer diffusion length 
(i.e.\/ $\kappa_S \ll \kappa_E$), the leading term is
stable and it is unstable if the diffusion length in the electrolyte is much
longer. The value of the ratio $\kappa_E/\kappa_S$ at which the sign changes
depends on the values of the other parameters. In the example of
Fig.~\ref{dopFDLAchifig} the critical ratio is $\kappa_E/\kappa_S=1.83$\/. 

For large $k\to \infty$, the dispersion relation has a limit 
\begin{equation}
\omega(k)=-F\,\Gamma\,{i^0_S}_\perp(0)\,\left(
\frac{\Psi^0_E(0)}{D_S}-\frac{\Psi^0_S(0)}{D_E}\right) + {\cal
O}(\frac{1}{k}).
\end{equation}
Like in the previous sections, the stability of small scale perturbations is
determined by the ratio of diffusion constants. They are stable if the
diffusivity in the semiconductor is much larger than in the electrolyte and
unstable if $D_S\ll D_E$\/. 
For the parameters in Fig.~\ref{dopFDLADfig}, the critical ratio
of diffusion constants is $D_S/D_E=0.90/2.0$\/.

With the boundary conditions discussed in this section, the current through
the interface is controlled by the ratio 
$\Psi^{\text{eq}}_S/\Psi^{\text{eq}}_E$\/. Like in the last section, short
wave length perturbations can be stabilized by increasing the current over a
certain threshold, i.e.\/ increasing the ratio of
$\Psi^{\text{eq}}_S/\Psi^{\text{eq}}_E$\/. In the example in 
Fig.~\ref{dopFDLApsifig} this value is $8.58/2.0$\/. 

Reducing the ratio $\Psi^{\text{eq}}_S/\Psi^{\text{eq}}_E$ to the
equilibrium value, i.e.\/ for $\Psi^{\text{eq}}_S\,\Psi^{\text{eq}}_E=\eta$,
the zero order current density $i_S^0(0)$ vanishes and the interface becomes
marginally stable. Below that value semiconductor material is deposited
and $\omega(k)$ changes sign for all $k$\/. 
In Fig.~\ref{dopFDLAi0fig} the growth rate of the flat interface $\dot{h}_0$
and the limit of the dispersion relation for large $k$ are plotted against
$\Psi^{\text{eq}}_S$ for the same parameters as in Fig.~\ref{dopFDLApsifig}, 
showing the two sign changes of $\omega(k\to\infty)$\/. The sign change at
$\Psi^{\text{eq}}_S=0.5$ is accompanied by a reversion of the growth
direction whereas at $\Psi^{\text{eq}}_S=8.58$ only the stability
properties change.

\section{Discussion}
\label{discussion}

Motivated by the discussion of electrochemical etching of silicon in HF
solutions  we developed a simplified model in
Sec.~\ref{stabilityanal}\/. This model has more than one field 
determining the interface motion,  includes a change of valence in the
dissolution reaction,  has nonlinear law of mass action type boundary
condition, and  includes a background reservoir for the reactants. The
transport in both, the semiconductor and the electrolyte is diffusive. We
performed a linear stability analysis of a flat interface for three limiting
cases of a simplified model and found two mechanisms that can cause a 
stabilization of the interface at high current densities. 

Firstly, a change of valence of the dissolution reaction with current density 
can stabilize the interface. The sign of the dispersion relation flips 
when the change of the valence with current density is large enough, namely
if ${i^0_S}_\perp(0)\,\frac{d\,\ln \nu(i)}{d\,i}|_{{i^0_S}_\perp(0)}>1$\/. 
This is true in
general, independent of the transport mechanisms in the semiconductor and
electrolyte, the number of reactants and reaction products, and the type of
boundary conditions. For electrochemical etching of silicon in hydrofluoric
acid the change of valence at the transition from pore formation at low
current densities to electropolishing at high current densities is large
enough. The stability of the interface at high current densities, after the
valence settled at the electropolishing value, cannot be explained with this
mechanism. However, there an oxide layer is formed which has to be dissolved 
chemically. At low current densities, when pores are formed, silicon is
dissolved directly. This oxide layer reduces the diffusivity in the
electrolyte considerably, leading to a stabilization of the interface. Our
model shows, that an interface propagating into a much more diffusive medium
is stable (given that the reactants are the rate limiting species, not the
reaction products). 

A second mechanism that at least partially stabilizes an interface are
nonlinear law of mass action type boundary conditions. While the sign of the
dispersion relation $\omega(k)$ at small $k$ is determined by the ratio of
diffusion lengths in semiconductor and electrolyte, the value at high $k$
as a function of the current density can change sign from positive to
negative. The sign of $\omega(k)$ for small $k$ is determined by the ratio
of diffusion lengths. The side with the larger diffusion length $1/\kappa$,
i.e.\/ the larger life time $\tau$, determines the stability of long wave
length perturbations. If the diffusion length of the  semiconductor
is much longer than the one of the electrolyte, the 
interface is unstable for long wavelengths. If
the diffusion length in the electrolyte is much larger, $\omega(k)$ is
negative for small $k$\/.

Our analysis shows, that a continuum model of surface growth, i.e.\/ a
moving boundary value problem for partial differential equations, can have a
transition from linear stability to instability with increasing current 
density. The two mechanisms discussed here are particularly interesting from
a theoretical point of view in that the effects of nonlinearities in the
model equations can still be handled analytically through
a linear stability analysis.

It is questionable whether the transition from porous silicon 
formation to electropolishing can be described by one of the discussed
mechanisms alone. The valence of the dissolution reaction of silicon does
change in this transition. But the implementation
in the model of Sec.~\ref{simplmodel} is oversimplified.
A more realistic model would include two alternative reaction pathways for
the species at the interface in the spirit of Eq.~(\ref{combinedmwg})\/. 
The analysis in this paper should be regarded as a stepping stone for 
the development of such models. 

\acknowledgments
M.~Rauscher thanks the DFG and the Bavarian government for support.
This work was supported by the Cornell Center for Materials Research
(CCMR), a Materials Research Science and Engineering Center of the
National Science Foundation (DMR-0079992).

\begin{figure}
\mycaption{
Sketch of the simplified model. The semiconductor is
above the interface $h(x,y)$ and the electrolyte below. Particles move to the
interface from both sides and react with each other, dissolving the
semiconductor.}
\label{setupfig} 
\end{figure}

\begin{figure}
\mycaption{
Dispersion relation in case of 
infinite life time and for
several values of $D_S$. 
Note the sign change at $D_S=50$\/.
The other parameters are set to
$\Gamma=1$, $D_E=1$, $d_E=10$, $d_S=20$, $\eta=1$, $I=-1$ and
$\Psi_{d_E}=20$\/. 
}
\label{doppelDLADfig}
\end{figure}

\begin{figure}
\mycaption{
Dispersion relation in case of infinite life time
and for various current densities. 
Note the sign change between $I=-1.8$ and $I=-1.95$\/.
The other parameters are
$\Gamma=1$, $D_E=1$, $D_S=0.5$, $d_E=10$, $d_S=20$, $\eta=1$, and
$\Psi_{d_E}=20$\/. 
}
\label{doppelDLAIfig} 
\end{figure}

\begin{figure}
\mycaption{
The dispersion relation in case of finite life time
in the electrolyte and infinite life time in the semiconductor for various
current densities. 
The sign of $\omega(k)$ for $k\to \infty$ changes at $I=-3.38$\/.
The other parameters are $F=1$, $\Gamma=1$, $D_E=10$,
$D_S=1$, $d_S=10$, $\eta=1$, $\kappa_E=1$, and $\Psi_E^{\text{eq}}=1$\/.}
\label{dopDLAFDLAIfig}
\end{figure}

\begin{figure}
\mycaption{
Dispersion relation in case of finite life time
in semiconductor and electrolyte for different values of $\kappa_E$\/. Note
that the sign of $\omega(k)$ for small $k$ changes at $\kappa_E=1.83$
 but not the sign for large $k$\/. 
The other parameters are 
$F=1$, $\Gamma=1$, $D_E=2$, $D_S=1$, $\kappa_S=1$,
$\epsilon=1$, $\Psi^{\text{eq}}_E=2$, and $\Psi^{\text{eq}}_S=10$\/.}
\label{dopFDLAchifig}
\end{figure}

\begin{figure}
\mycaption{
Dispersion relation in case of finite life time in
semiconductor and electrolyte for different values of $D_S$\/. Note that the
limit of $\omega(k)$ for $k\to\infty$ changes sign at
$D_S=0.90$\/. The other parameters are 
$F=1$, $\Gamma=1$, $D_E=2$, $\kappa_S=1$, $\kappa_E=10$,
$\epsilon=1$, $\Psi^{\text{eq}}_E=2$, and $\Psi^{\text{eq}}_S=10$\/.}
\label{dopFDLADfig}
\end{figure}

\begin{figure}
\mycaption{
Dispersion relation in case of finite life time in
semiconductor and electrolyte for different values of $\Psi^{\text{eq}}_S$,
i.e.\/ different current densities. Note that $\omega(k)$ for $k\ll 1$ 
remains positive. The other parameters are
$F=1$, $\Gamma=1$, $D_E=2$, $D_S=1$, $\kappa_S=1$, $\kappa_E=10$,
$\epsilon=1$, and $\Psi^{\text{eq}}_E=2$\/.}
\label{dopFDLApsifig}
\end{figure}

\begin{figure}
\mycaption{
The growth velocity of the flat interface
(dashed line) and the limiting value of $\omega(k)$ for $k\to\infty$ (dotted
line) are plotted against
$\Psi^{\text{eq}}_S$\/. The growth velocity and $\omega(\infty)$ change 
sign at $\Psi^{\text{eq}}_S=0.5$\/. At $\Psi^{\text{eq}}_S=8.58$ only
$\omega(\infty)$ changes sign. The other parameters are the same as in
Fig.~\protect\ref{dopFDLApsifig}.}
\label{dopFDLAi0fig}
\end{figure}
\end{multicols}
\widetext
\pagestyle{empty}
\eject
\huge

\begin{center}
\epsfig{file=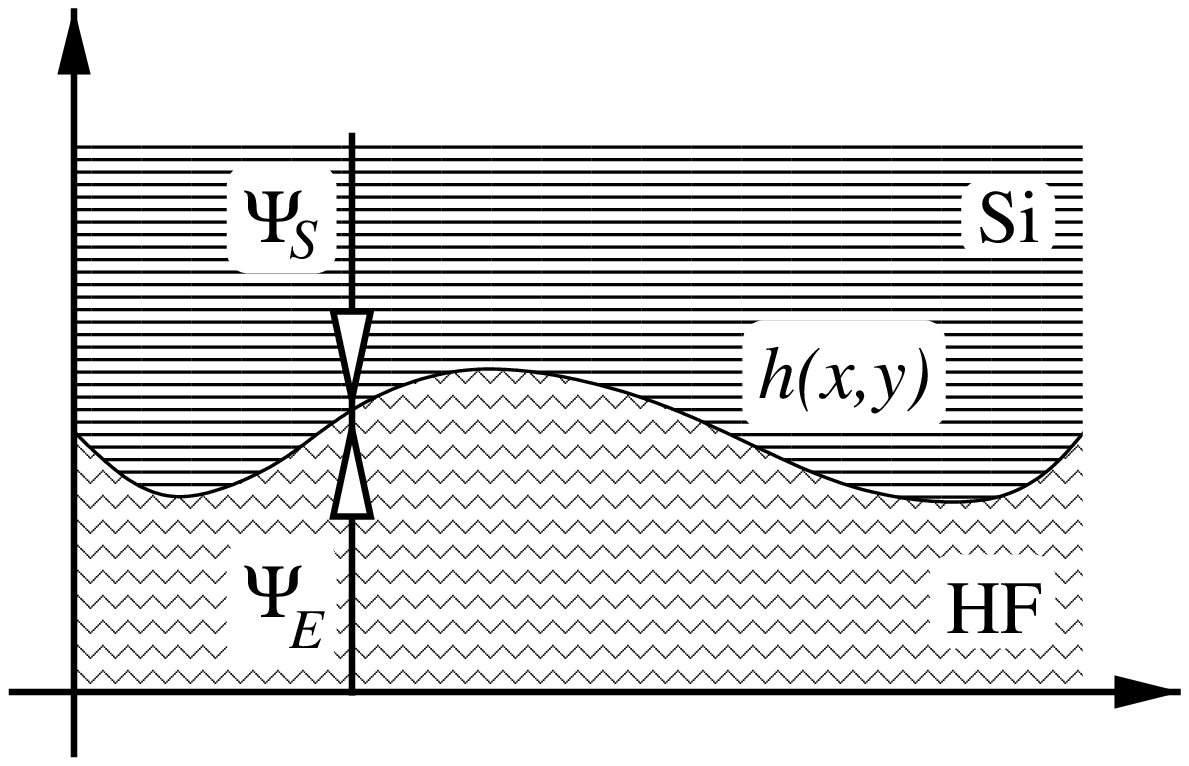,width=\linewidth}
\end{center}
\vfill

Fig.~\ref{setupfig} 
\eject

\begin{center}
\epsfig{file=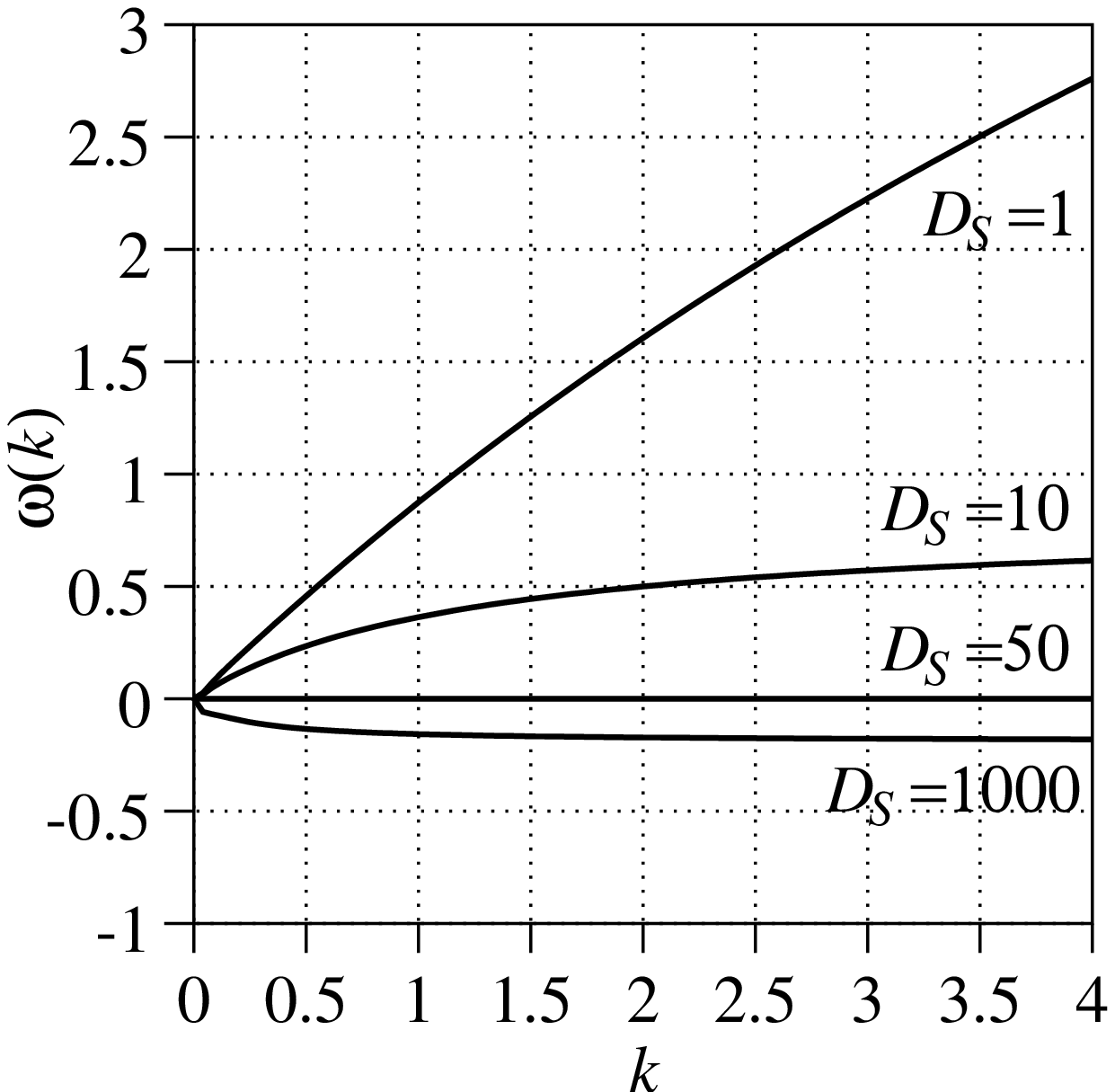,width=\linewidth}
\end{center}
\vfill

Fig.~\ref{doppelDLADfig}
\eject

\begin{center}
\epsfig{file=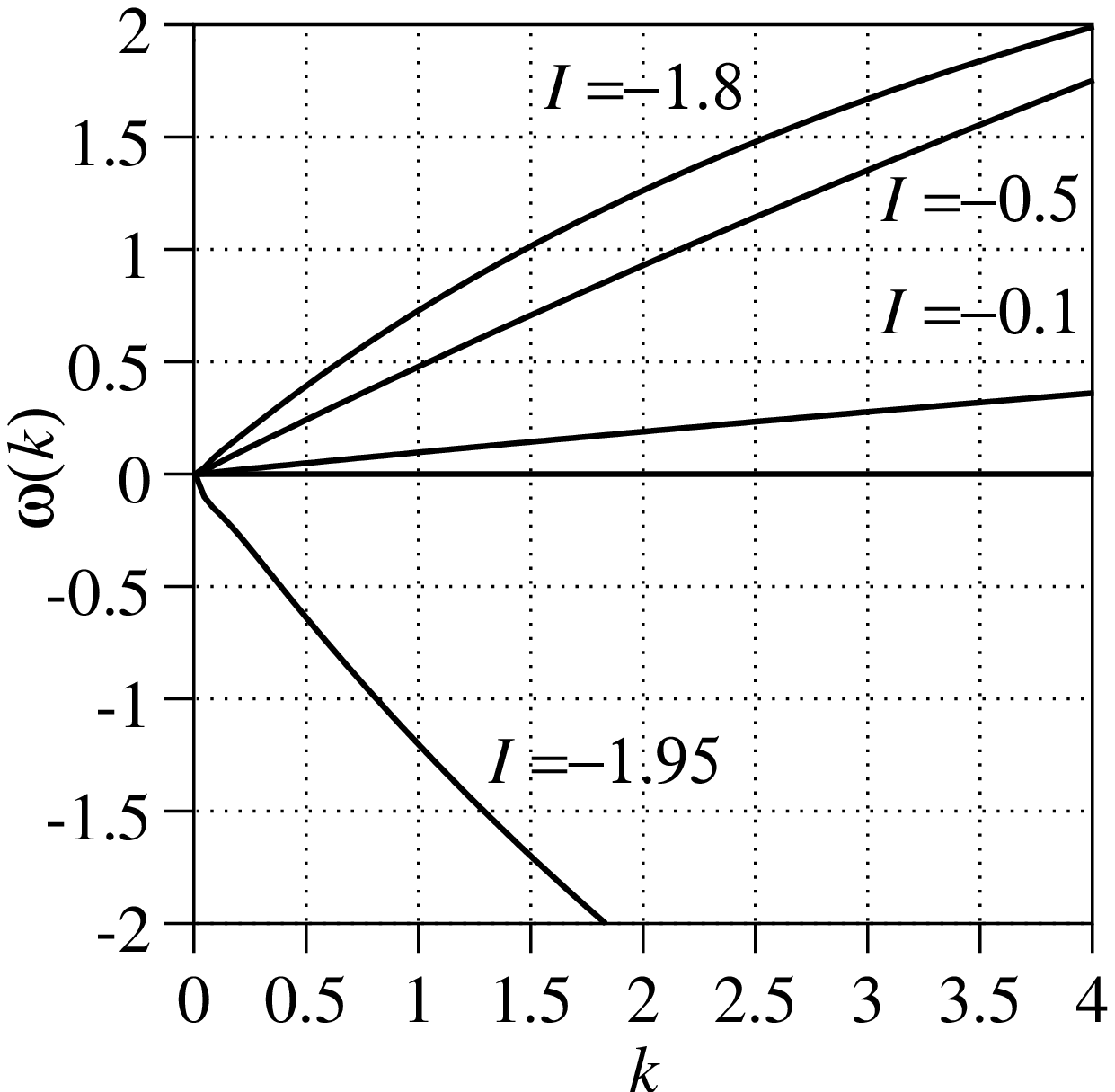,width=\linewidth}
\end{center}
\vfill

Fig.~\ref{doppelDLAIfig} 
\eject

\begin{center}
\epsfig{file=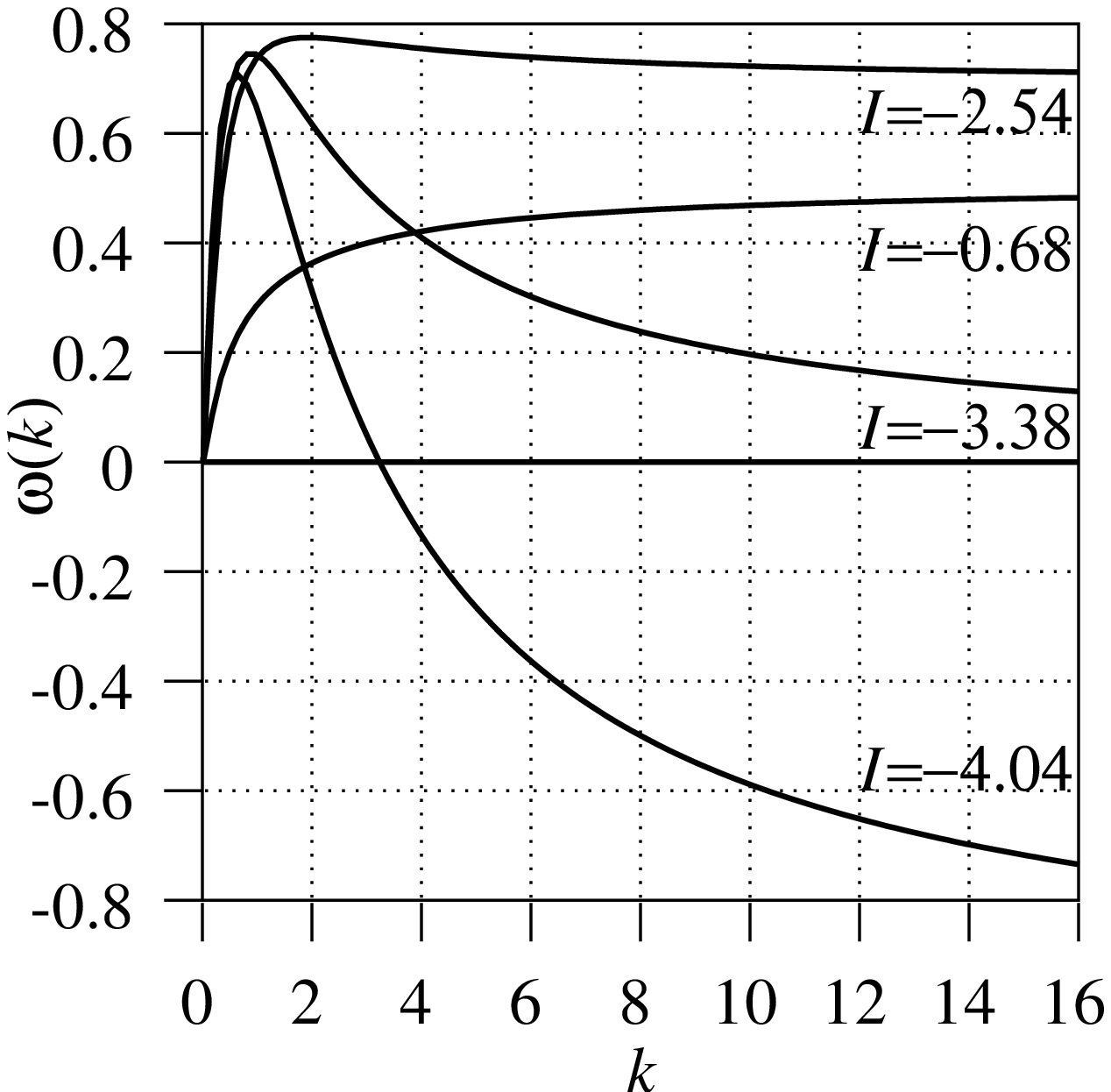, width=\linewidth}
\end{center}
\vfill

Fig.~\ref{dopDLAFDLAIfig}
\eject

\begin{center}
\epsfig{file=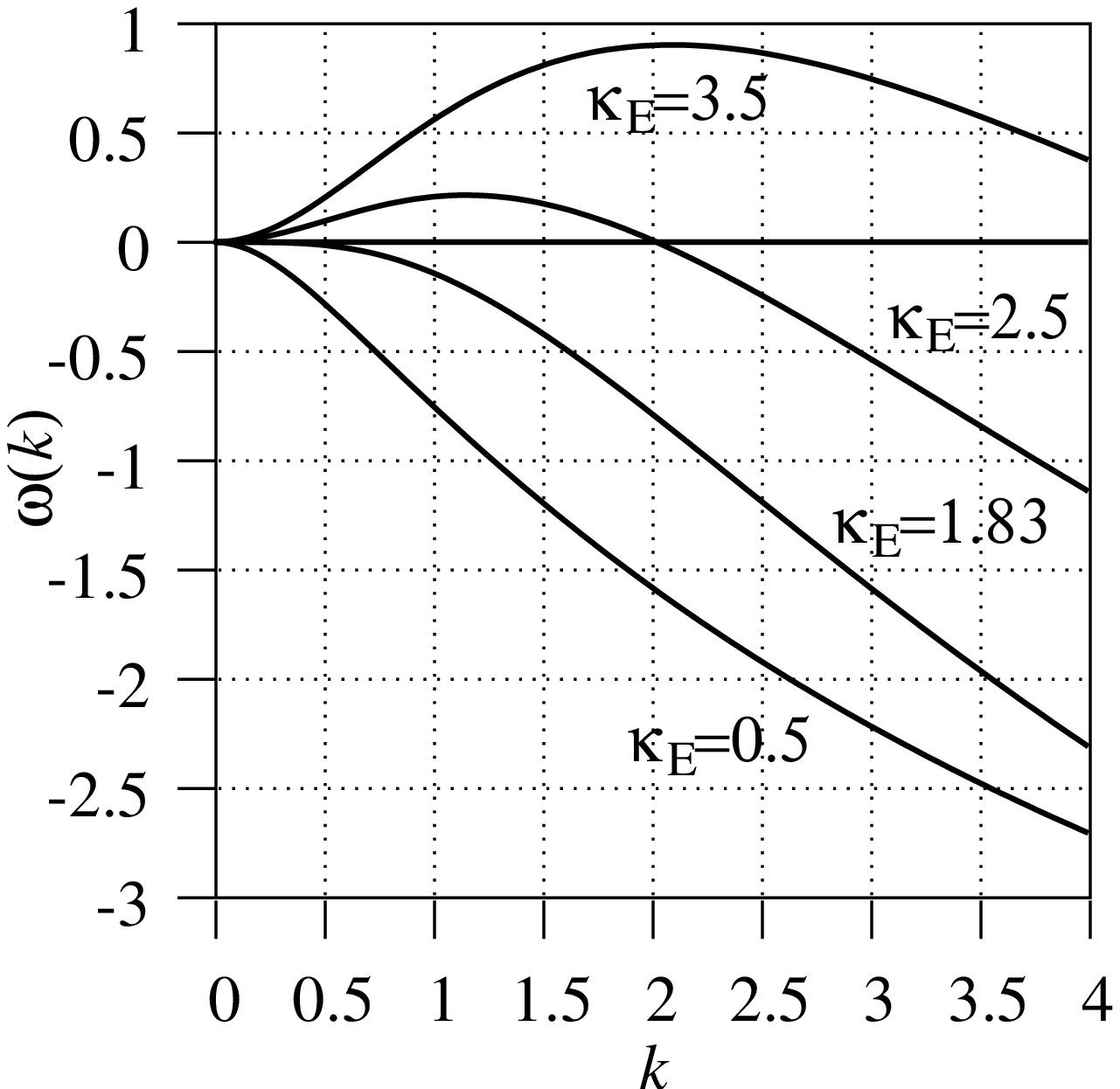, width=\linewidth}
\end{center}
\vfill

Fig.~\ref{dopFDLAchifig}
\eject

\begin{center}
\epsfig{file=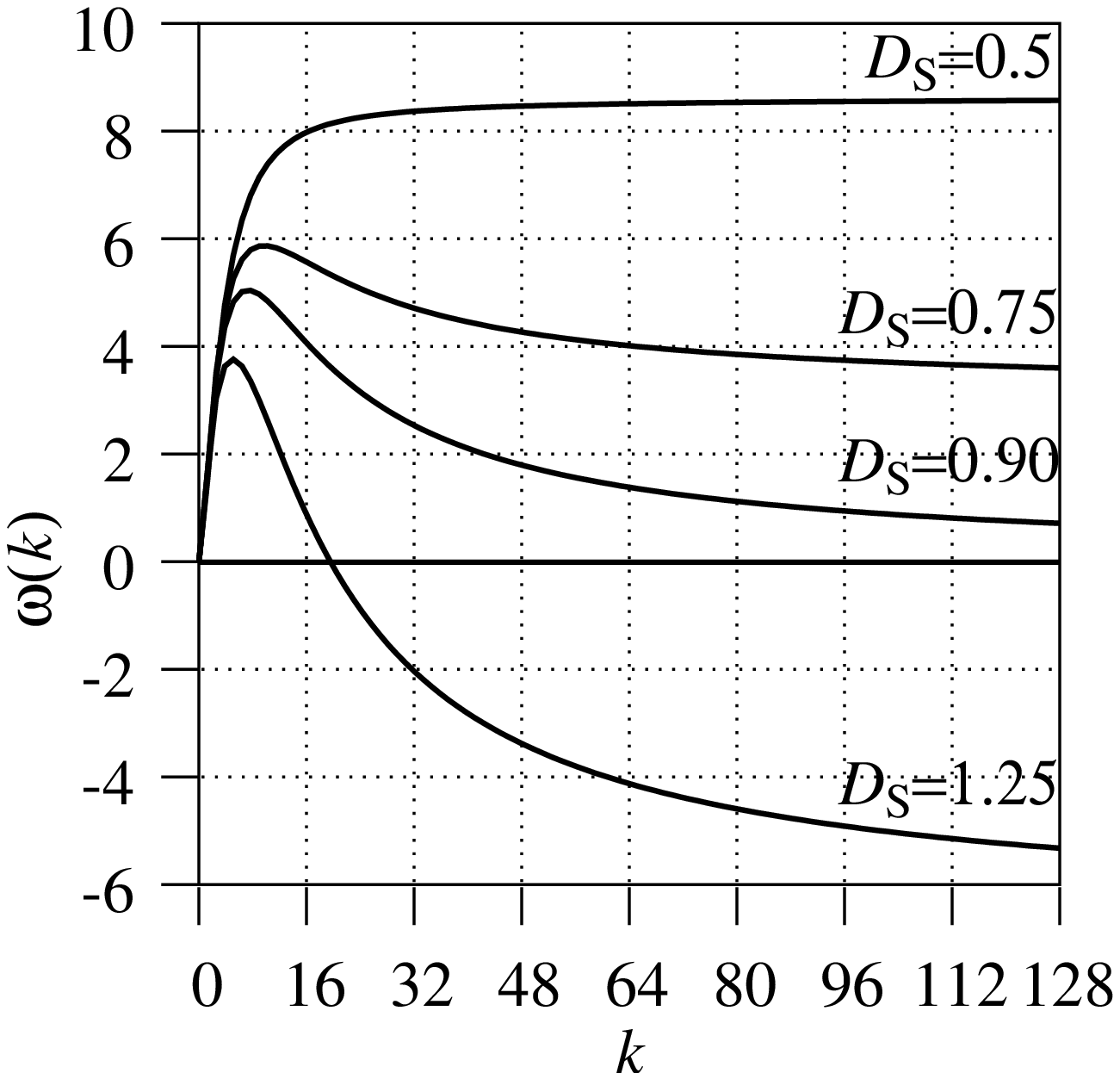,width=\linewidth}
\end{center}
\vfill

Fig.~\ref{dopFDLADfig}
\eject

\begin{center}
\epsfig{file=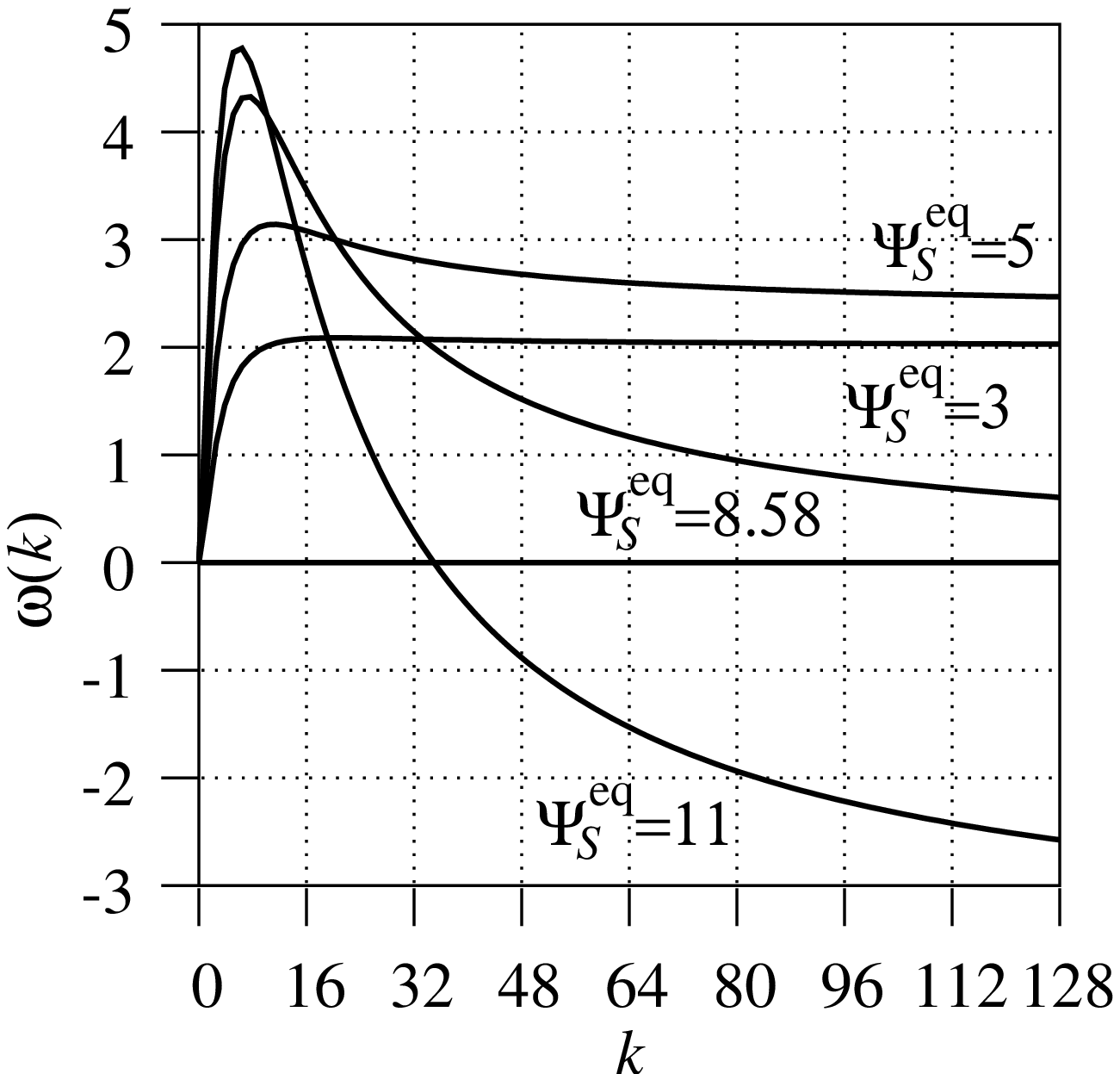,width=\linewidth}
\end{center}
\vfill

Fig.~\ref{dopFDLApsifig}
\eject

\begin{center}
\epsfig{file=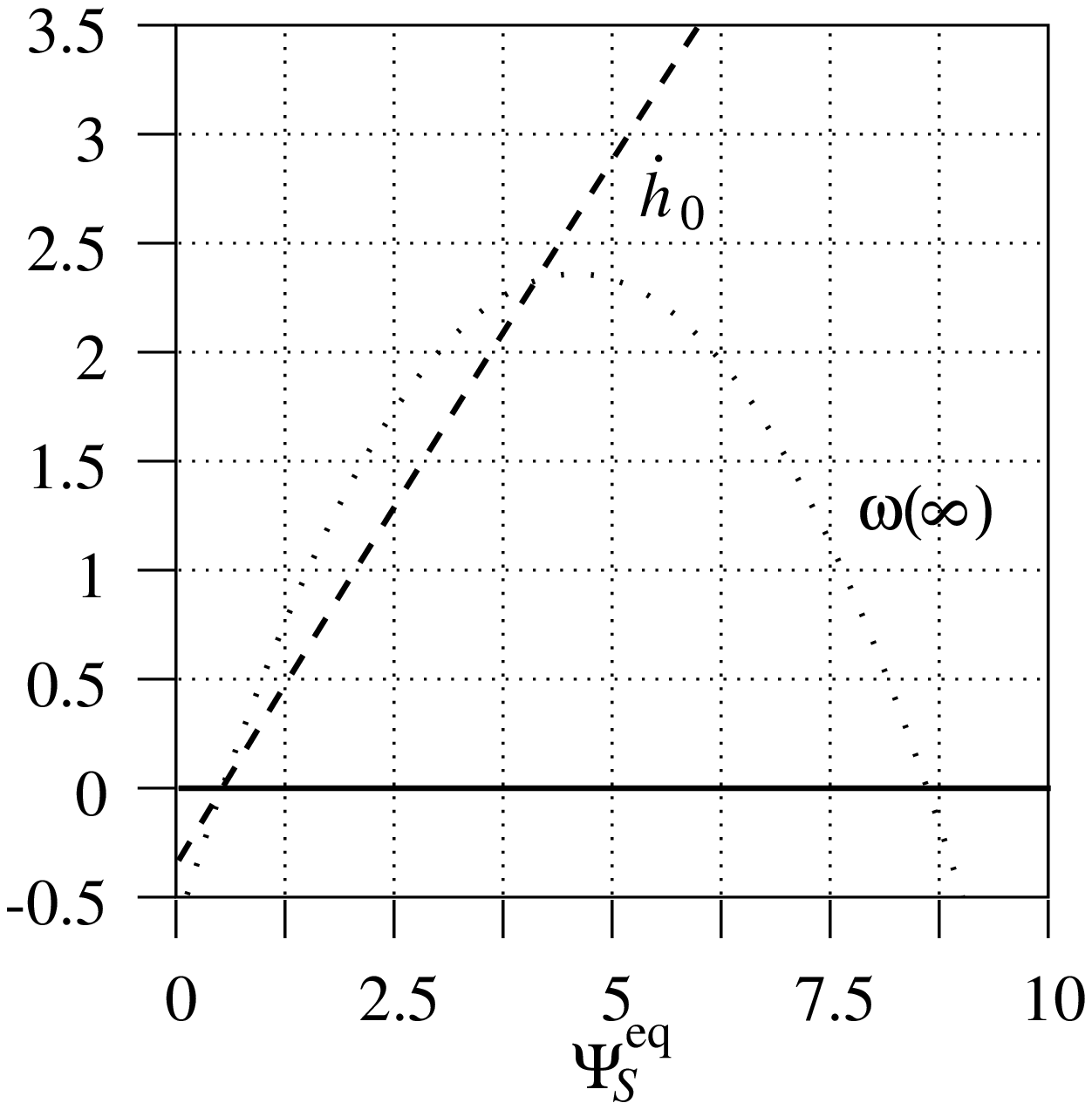,width=\linewidth}
\end{center}
\vfill

Fig.~\ref{dopFDLAi0fig}

\begin{thebibliography}{10}
\bibitem{ulhir} A.~Ulhir, Bell System Tech.\ J. {\bf 35}, 333 (1956)
\bibitem{turner} D.~R.~Turner, J.~Electrochem.\ Soc. {\bf 105}, 402 (1958)
\bibitem{canham} L.~T.~Canham, Appl.\ Phys.\ Lett.~{\bf 57}, 1046 (1990)
\bibitem{lehmann} V.~Lehmann and U.~G\"{o}sele, Appl.\ Phys.\ Lett.~{\bf 58},
856 (1991)
\bibitem{parkhutik99} V.~Parkhutik, Solid-State Electron.~{\bf 43}, 1121 (1999)
\bibitem{john} G.~C.~John and V.~A.~Singh, Phys.\ Rep.~{\bf
263}, 93 (1995)
\bibitem{erlebacher94} J.~Erlebacher, K.~Sieradzki, and P.~C.~Searson,
J.~Appl.\/ Phys.\/ {\bf 76}, 182 (1994)
\bibitem{collins89} R.~L.~Smith and S.~D.~Collins, Phys.\/ Rev.~A {\bf 39}, 5409
(1989)
\bibitem{parkhutik93} V.~P.~Parkuthik {\it et al.}, Appl.\/
Phys.\/ Lett.\/ {\bf 62}, 366 (1993)
\bibitem{kang93} Y.~Kang and J.~Jorn\'{e}, J.~Electrochem.\/ Soc.\/ {\bf 140},
2258 (1993)
\bibitem{valance97} A.~Valance, Phys.\/ Rev.~B {\bf 55}, 9706 (1997)
\bibitem{foell00} H.~F\"{o}ll, J.~Carstensen, M.~Christophersen, and G.~Hasse, 
phys.\/ stat.\/ sol.~(a) {\bf 182}, 7 (2000)
\bibitem{gerischer} H.~Gerischer, P.~Allongue, and V.~Costa Kieling, 
Ber.\ Bunsenges.\ Phys.\ Chem.\ {\bf 97}, 753 (1993)
\bibitem{lehmanndiss} V.~Lehmann, Ph.D.~thesis, 
Friedrich-Alexander-Universit\"{a}t Erlangen-N\"{u}rnberg, 1988
\bibitem{cercignani88} C.~Cercignani, {\it The Boltzmann Equation and Its
Applications}, Applied
Mathematical Science Volume {\bf 67} (Springer, New York, 1988)
\bibitem{bockris} J.~O'M.~Bockris and A.~K.~N.~Reddy, {\it Modern 
Electrochemistry} (Plenum Press, New York, 1970), Vol.~1 and 2
\bibitem{henisch} H.~K.~Henisch, {\it Semiconductor Contacts} (Clarendon,
Oxford, 1985)
\bibitem{remarkonunits} In the literature on physical chemistry, $K_w$ and 
$K_{\text{HF}}$ are dimensionless, since not the concentrations but the 
dimensionless activity defines the equilibrium constants. However, in linear 
approximation, the chemical activity is proportional to the concentration. 
\bibitem{atkins86} P.~W.~Atkins, {\it Physical Chemistry} (Oxford
University Press, Oxford, 1986/87), 3.\/ edition
\bibitem{pelce} P.~Pelc\'{e} in {\it Dynamics of Curved Fronts}, edited by P.~Pelc\'{e}
(Academic Press, London 1988), part I
\bibitem{vanmaekelbergh} D.~Vanmaekelbergh and P.~C.~Searson, J.~Electrochem.\
Soc.\ {\bf 141}, 697 (1994)
\bibitem{rauscher00} M.~Rauscher and H.~Spohn,
J.~Por.~Mater. {\bf 7}, 345 (2000)
\bibitem{voltage} 
If $\Psi_{E/S}$ is interpreted as the electrostatic potential 
$V$ (or as $-V$),  $\sqrt{D_{E/S}\,\tau_{E/S}}$ is the electrostatic screening length.
\bibitem{lehmann93} V.~Lehmann, J.~Electrochem.\/ Soc.\/ {\bf 140}, 2836
(1993)
\bibitem{DLEXX} J.~Krug and  P.~Meakin, 
Phys.\/ Rev.\/ Lett.\/ {\bf 66}, 703 (1991)
\end{thebibliography}
\end{document}